\documentclass[english,prl,twocolumn,showpacs]{revtex4}
\usepackage[OT1]{fontenc}
\usepackage[latin1]{inputenc}
\usepackage{graphicx}
\usepackage{amssymb}

\makeatletter

\providecommand{\boldsymbol}[1]{\mbox{\boldmath $#1$}}

\makeatletter

\makeatletter
\makeatletter
\usepackage{latexsym}

\usepackage{bm}

\makeatother
\makeatother

\makeatother

\usepackage{babel}
\makeatother
\begin{document}

\title{Interplay of polarization geometry and rotational dynamics in high
harmonic generation from coherently rotating linear molecules}

\author{F.H.M. Faisal$^{1,2}$}

\author{A. Abdurrouf$^{1}$}

\affiliation{$^{1}$Fakultät für Physik, Universität Bielefeld, Postfach 100131,
D-33501 Bielefeld, Germany}

\affiliation{$^{2}$ITAMP, Harvard-Smithsonian Center for Astrophysics, 60 Garden
St., Cambridge, MA 02138, USA}

\begin{abstract}
Recent reports on intense-field pump-probe experiments for high harmonic
generation from coherently rotating linear molecules, have revealed
remarkable characteristic effects of the simultaneous variation of
the polarization geometry and the time delay on the high harmonic
signals. We analyze the effects and give a unified theoretical account
of the experimental observations. 
\end{abstract}


\maketitle
The phenomenon of high harmonic generation (HHG) from atoms or molecules
in intense laser fields can be thought of as a {}``fusion'' of $n$
laser photons, each of energy $\hbar\omega$, into a \textit{single}
harmonic photon of an enhanced energy $\hbar\Omega=n\hbar\omega$.
This might seem surprising at first since the photons do \textit{not}
interact with each other and therefore can not {}``fuse'' on their
own. However, a bound electron interacting with a laser pulse can
absorb $n$ photons from the laser field, go into highly excited virtual
states and can return to the same bound state by releasing precisely
the excitation energy ($n\hbar\omega$) as a single harmonic photon.
Note that at the end of the coherent process the electron does not
change its state -- it merely acts as a {}``catalyst'' of the process.
The phenomenon is currently being vigorously investigated, both experimentally
and theoretically, specially in connection with dynamic alignments
of molecules (e.g. \cite{exp_the}).

\begin{figure}
\begin{centering}
\includegraphics[scale=0.5]{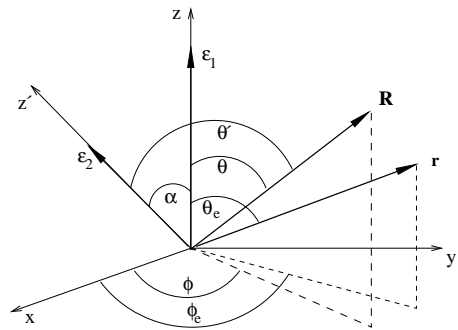} 
\par\end{centering}

\caption{\label{fig:01} A schematic diagram defining: molecular axis, $\boldsymbol{R}$,
electron position $\boldsymbol{r}$, pump polarization $\boldsymbol{\epsilon}_{1}$,
probe polarization $\boldsymbol{\epsilon}_{2}$; z and z' axes lie
on common z-z'-x plane; fields propagate along y-axis.}
\end{figure}

Recently a number of remarkable pump-probe experiments for high harmonic
generation with intense femtosecond laser pulses from coherently rotating
linear molecules (e.g. N$_{2}$, O$_{2}$, CO$_{2}$, HC$\equiv$CH)
have been reported in this journal and elsewhere (e.g. \cite{ita-02,miy-01,kan-01,miy-02,kaj-02,tor-01}).
These experiments measure the HHG signals as a function of the delay-time,
$t_{d}$, between a pump pulse that sets the molecule in coherent
rotation, and a probe pulse that generates the high harmonic signal
from the rotating molecule. The changes in the dynamic signals are
then recorded by varying the angle, $\alpha$, between the polarizations
of the two pulses. Fig. \ref{fig:01} shows a schematic diagram of
the various vectors involved in the pump-probe experiments. The geometric
angle $\alpha$ is the operational angle in the laboratory, although
at times it is erroneously identified with the angle $\theta$ (or
$\theta'$); the latter is a quantum variable, not measured in these
experiments. Here we derive an explicit theoretical expression for
the HHG signal as a simultaneous function of the geometric angle $\alpha$
and the delay-time $t_{d}$ and analyze the experimental observations.
The results provide a unified theoretical account of the observed
effects. %

Let the total Hamiltonian of the molecular system interacting with
a pump pulse $L_{1}$ at a time $t$, and a probe pulse $L_{2}$ applied
after a delay-time $t_{d}$, be written, within the Born-Oppenheimer
approximation as (e.g. \cite{fai_rou-01,rou_fai-01}): \begin{equation}
H_{tot}(t)=H_{N}^{(0)}+V_{N-L_{1}}(t)+H_{e}^{(0)}+V_{e-L_{2}}(t-t_{d})\label{HamTot}\end{equation}
where the subscripts $N$ and $e$ stand for the nuclear and the electronic
subsystems, respectively. An intense femtosecond pump-pulse is assumed
to interact with the molecular polarizability, via $V_{N-L_{1}}(t)$,
and sets it into coherent free rotation.

The coherent rotational motion \cite{early_works} is described by
the nuclear wavepacket states created by the pump pulse: \begin{equation}
\left|\Phi_{J_{0}M_{0}}(t)\right\rangle =\sum_{J}C_{JM}^{J_{0}M_{0}}(t)e^{-\frac{i}{\hbar}E_{JM}t}\left|JM\right\rangle .\label{IniWavFun}\end{equation}
 Each wavepacket state (\ref{IniWavFun}) evolves \textit{one-to-one}
from an initially occupied ensemble of eigen states, $\left|J_{0}M_{0}\right\rangle $,
populated with a Boltzmann distribution $\rho(J_{0})=\frac{1}{Z_{P}}e^{-E_{J_{0}M_{0}}/kT}$,
where $Z_{P}$ is the partition function. Thus, after the pump pulse,
the initial state of the molecule is characterized by the ensemble
of product states, $\left|\chi_{i}(t)\right\rangle $, with $i\equiv\left\{ e,J_{0},M_{0}\right\} $,
composed of the ground electronic state $\left|\phi_{e}^{(0)}(t)\right\rangle $
and the coherent wavepackets $\left|\Phi_{J_{0}M_{0}}(t)\right\rangle $:
\begin{equation}
\left|\chi_{i}(t)\right\rangle =\left|\phi_{e}^{(0)}(t)\right\rangle \left|\Phi_{J_{0}M_{0}}(t)\right\rangle .\label{IniSta}\end{equation}
 Generalizing the well-known strong-field KFR (Keldysh-Faisal-Reiss)
approximation (e.g. \cite{bec-01}) to the present molecular case,
we write the wavefunction of the system, evolving from each of the
ensemble of the initial states (\ref{IniSta}) as:\begin{equation}
\left|\Psi(t)\right\rangle =\left|\chi_{i}(t)\right\rangle +\int dt'G_{0}(t,t')V_{e-L2}(t'-t_{d})\left|\chi_{i}(t')\right\rangle \label{KFRWavFun}\end{equation}
 where, the Green's function $G_{0}(t,t')$ of the system is given
by \begin{eqnarray}
G_{0}(t,t') & = & -\frac{i}{\hbar}\theta(t-t')\sum_{j\boldsymbol{p}JM}\left|\phi_{j}^{(+)}\right\rangle \left|\phi_{\boldsymbol{p}}(t-t_{d})\right\rangle \nonumber \\
 &  & \times\left|\Phi_{JM}(t)\right\rangle e^{-\frac{i}{\hbar}E_{j}^{+}(t-t')}\left\langle \Phi_{JM}(t')\right|\nonumber \\
 &  & \times\left\langle \phi_{\boldsymbol{p}}(t'-t_{d})\right|\left\langle \phi_{j}^{(+)}\right|\label{TotGreFun}\end{eqnarray}
 where, $\left|\phi_{j}^{(+)}\right\rangle $ are ionic orbitals and
$\left|\phi_{\boldsymbol{p}}(t)\right\rangle $ are Volkov states
(e.g. \cite{bec-01}).%
\begin{figure}
\begin{centering}
\includegraphics[scale=0.3]{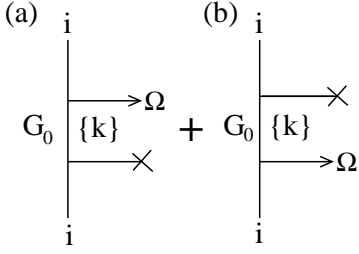} 
\par\end{centering}

\caption{\label{fig:02} Quantum amplitude for coherent emission of a high
harmonic photon (frequency $\Omega$) is the sum of two diagrams,
(a) direct, and (b) time-reversed; probe-interaction (line-$\times$),
photon emission (arrow); intermediate propagators, $G_{0}$; Volkov
wave-vector $\bm{k}$; $i\equiv$ Eq.(\ref{IniSta}). }
\end{figure}

The quantum transition amplitude for the coherent emission of a harmonic
photon of energy $\hbar\Omega=n\hbar\omega$, from an initial state
(\ref{IniSta}) evolving into (\ref{KFRWavFun}) and recombining back
into the same state (\ref{IniSta}), is given by the sum of a `direct'
and a `time reversed' diagram for the photon emission process (cf.
Fig. \ref{fig:02}). Writing out the amplitude analytically using
Eqs. (\ref{IniWavFun}) to (\ref{TotGreFun}), assuming the {}``adiabatic
nuclei'' condition, $Max(\Delta E_{J,J'})\ll E_{e}$, carrying out
the lengthy algebra, and modulo-squaring the result, we obtain the
coherent HHG emission probability for each initial state (\ref{IniSta}).
Taking the statistical average of the independent probabilities for
the ensemble of initial states (\ref{IniSta}), we obtain the scaled
HHG signal {}``per molecule'', as an explicit function of $\alpha$
and $t_{d}$: \begin{eqnarray}
S^{(n)}(t_{d},\alpha) & = & \sum_{J_{0}M_{0}}\rho(J_{0})\left|\left\langle \Phi_{J_{0}M_{0}}(t_{d})\right|T^{(n)}(\theta,\phi;\alpha)\right.\nonumber \\
 &  & \times\left.\left|\Phi_{J_{0}M_{0}}(t_{d})\right\rangle \vphantom{T^{(n)}}\right|^{2}\label{Sig}\end{eqnarray}
 where \begin{eqnarray}
T^{(n)}(\theta,\phi;\alpha) & = & \sum_{L,M,l,l'}a_{z'z'}^{(n)}(l,l',L;m)\frac{4\pi}{2L+1}Y_{LM}(\alpha,0)\nonumber \\
 &  & \times Y_{LM}(\theta,\phi)\label{HhgOpe}\end{eqnarray}
 with, $L=(|l-l'|,(l+l'))$, $M=(-L,L)$; the parameters $a_{z'z'}^{(n)}(l,l',L;m)$
are given by rather lengthy but explicit expressions \cite{rou_fai-01}
that depend on the partial angular momenta $l(l')$ of the active
electron and their conserved projection, $m$, on the molecular axis,
on the matrix elements of the absorption and recombination transition-dipoles,
and on the usual vector addition coefficients.

Specializing Eq. (\ref{HhgOpe}) to the case of N$_{2}$ (molecular
orbital symmetry $\sigma_{g}$, $m=0$; dominant $l(l')=0,2,4$),
we get, in an ordinary trigonometric representation, an analytic expression
of the dynamic HHG signal for $N_{2}$: \begin{eqnarray}
S^{(n)}\left(t_{d},\alpha\right) & = & p_{1}+p_{2}\left\langle \left\langle \cos^{2}\theta'\right\rangle \right\rangle \left(t_{d}\right)\nonumber \\
 &  & +p_{3}\left\langle \left\langle \cos^{2}\theta'\right\rangle \left\langle \cos^{2}\theta'\right\rangle \right\rangle \left(t_{d}\right)\nonumber \\
 &  & +p_{4}\left\langle \left\langle \cos^{4}\theta'\right\rangle \right\rangle \left(t_{d}\right)+\cdots\nonumber \\
 &  & +p_{10}\left\langle \left\langle \cos^{6}\theta'\right\rangle \left\langle \cos^{6}\theta'\right\rangle \right\rangle \left(t_{d}\right)\label{NitHhgSig}\end{eqnarray}
 where $\cos\theta'=\cos\alpha\cos\theta+\sin\alpha\sin\theta\cos\phi$.
Similarly, for O$_{2}$, ($\pi_{g}$ symmetry, $m=1$, and dominant
$l(l')=2,4$) we get, \begin{eqnarray}
S\left(t_{d},\alpha\right) & = & q_{1}\left\langle \left\langle \sin^{2}\theta'\cos^{2}\theta'\right\rangle ^{2}\right\rangle \left(t_{d}\right)\nonumber \\
 &  & +q_{2}\left\langle \left\langle \sin^{2}\theta'\cos^{2}\theta'\right\rangle \vphantom{\left\langle \sin^{2}\theta'\cos^{2}\theta'\right\rangle ^{2}}\right.\nonumber \\
 &  & \times\left.\vphantom{\left\langle \sin^{2}\theta'\cos^{2}\theta'\right\rangle ^{2}}\left\langle \sin^{2}\theta'\cos^{4}\theta'\right\rangle \right\rangle \left(t_{d}\right)\nonumber \\
 &  & +\cdots+q_{6}\left\langle \left\langle \sin^{2}\theta'\cos^{6}\theta'\right\rangle ^{2}\right\rangle \left(t_{d}\right).\label{OxyHhgSig}\end{eqnarray}
 The coefficients $p$'s and $q$'s are determined by simple combinations
of the parameters $a_{z'z'}^{(n)}(l,l',L;m)$ \cite{rou_fai-01}.
We note that for $\alpha=$0, Eqs. (\ref{NitHhgSig}) and (\ref{OxyHhgSig})
reduce correctly to the special limits \cite{fai_rou-01}.

\begin{figure}[b]

\begin{centering}
\includegraphics[scale=0.3]{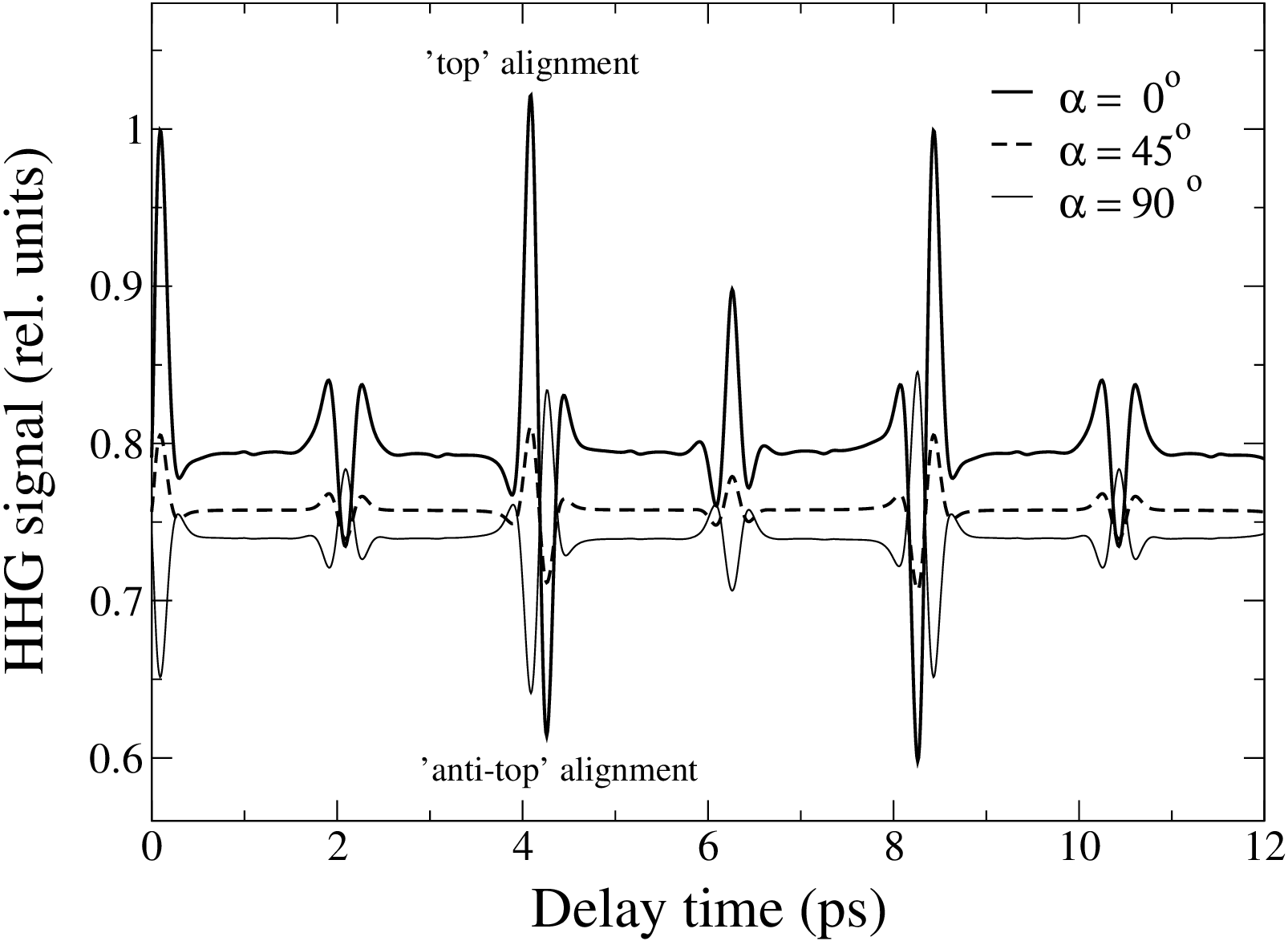} 
\par\end{centering}

\caption{\label{fig:03} Calculated $19$th harmonic dynamic signal for N$_{2}$
for various pump-probe polarization angles, i.e. $\alpha=0^{o}$,
$\alpha=45^{o}$, and $\alpha=90^{o}$; pump intensity $I=0.8\times10^{14}$W/cm$^{2}$,
probe intensity, $I=1.7\times10^{14}$ W/cm$^{2}$, duration $40\,\mathrm{fs}$,
and wavelength $800\,\mathrm{nm}$; Boltzmann temperature 200 K.}
\end{figure}

In Fig. \ref{fig:03} we show the results of computations using Eq.
(\ref{NitHhgSig}) for the dynamic signals from N$_{2}$ as a function
of the delay time $t_{d}$, at three different relative polarization
angles, $\alpha=$0$^{o},\,45^{o}$, and $90^{o}$. The results show
the full revival with a period $T_{rev}\equiv\frac{1}{2Bc}=8.4\,\mathrm{ps}$,
and the fractional $\frac{1}{2}$ and $\frac{1}{4}$ revivals, for
all three $\alpha$ values. The $\left\langle \left\langle \cos^{2}\theta\right\rangle \right\rangle \left(t_{d}\right)$
term is known to govern the $\frac{1}{2}$ and $\frac{1}{4}$ revivals
and the associated Raman allowed spectral lines (e.g.\cite{exp_the,fai_rou-01}).
Remarkably, the signals for $\alpha=$0$^{o}$ and $\alpha=90^{o}$
are found to be in opposite phase, while that for $\alpha=0^{o}$
and $\alpha=45^{o}$ are in the same phase. Exactly the same phase
relation between the $\alpha$-dependence of the $t_{d}$-signal from
$N_{2}$ has been observed in recent experiments (e.g. \cite{miy-01,kan-01,miy-02}).

To analyze their origin, we consider the leading term of Eq. (\ref{NitHhgSig})
for N$_{2}$, more explicitly. (Below, we omit the argument $\left(t_{d}\right)$
for the sake of brevity.) Noting that $\left\langle \cos^{2}\theta'\right\rangle =\frac{1}{2}\sin^{2}\alpha+(\cos^{2}\alpha-\frac{1}{2}\sin^{2}\alpha)\left\langle \cos^{2}\theta\right\rangle $
we get: \begin{eqnarray}
S\left(t_{d};\alpha\right) & \approx & \left(p_{1}+p_{2}\frac{1}{2}\sin^{2}\alpha\right)+p_{2}\left(\cos^{2}\alpha-\frac{1}{2}\sin^{2}\alpha\right)\nonumber \\
 &  & \times\left\langle \left\langle \cos^{2}\theta\right\rangle \right\rangle +\cdots\label{NitSigApp}\end{eqnarray}
 Therefore, for the parallel polarizations we have, $S\left(t_{d};0^{o}\right)\approx p_{1}+p_{2}\left\langle \left\langle \cos^{2}\theta\right\rangle \right\rangle $
and for the perpendicular polarization, $S\left(t_{d};90^{o}\right)\approx\left(p_{1}+\frac{1}{2}p_{2}\right)-\frac{1}{2}p_{2}\left\langle \left\langle \cos^{2}\theta\right\rangle \right\rangle $.
Clearly due to the opposite sign of the $\left\langle \left\langle \cos^{2}\theta\right\rangle \right\rangle $
term, they vary in \textit{opposite} phase to each other from their
respective bases. In contrast, the signal at $\alpha=45^{o}$, $S\left(t_{d},45^{o}\right)\approx\left(p_{1}+\frac{1}{4}p_{2}\right)+\frac{1}{4}p_{2}\left\langle \left\langle \cos^{2}\theta\right\rangle \right\rangle $,
has the same sign of the $\left\langle \left\langle \cos^{2}\theta\right\rangle \right\rangle $
term as for $\alpha=$0, which makes them to vary \textit{in} phase.
These behaviors are what can be seen in the full calculations in Fig.
3, and they also agree with the recent experimental observations (e.g.
\cite{ita-02,miy-01,kan-01,miy-02}). The simple formula (\ref{NitSigApp})
predicts further that the extrema of the signal should occur for $\sin\alpha\cos\alpha=0$,
with a maximum at $\alpha=$0$^{o}$ and a minimum at $\alpha=90^{o}$.
This is also what has been seen experimentally \cite{ita-02,miy-02}.
Finally, (\ref{NitSigApp}) predicts a {}``magic angle'' $\alpha_{c}=\arcsin\sqrt{\frac{2}{3}}\approx54.7^{o}$,
given by the condition $(\cos^{2}\alpha_{c}-\frac{1}{2}\sin^{2}\alpha_{c})=0$
at which the HHG signals become essentially independent of the delay
$t_{d}$ between the pulses. Exactly such a {}``magic'' crossing
angle for N$_{2}$ signals has been observed experimentally \cite{miy-02}.
We may point out that this geometry can be used in femtosecond pulse-probe
experiments to generate an essentially steady HHG signal from freely
rotating N$_{2}$.%
\begin{figure}
\begin{centering}
\includegraphics[scale=0.3]{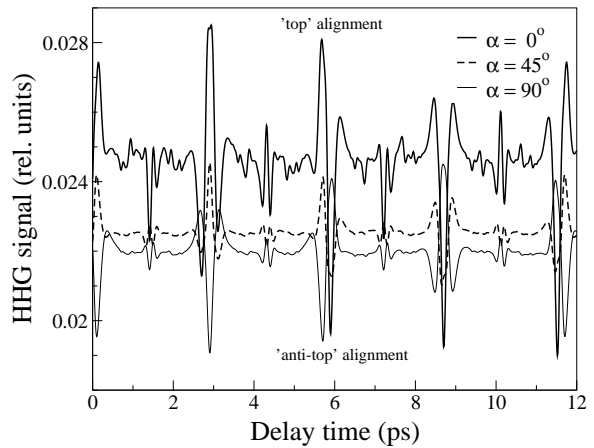} 
\par\end{centering}

\caption{\label{fig:04}Calculated $19th$ HHG spectrum of O$_{2}$ for various
pump-probe polarizations angle, i.e. $\alpha=0^{o}$, $\alpha=45^{o}$,
and $\alpha=90^{o}$; pump intensity $I=0.5\times10^{14}$W/cm$^{2}$,
probe intensity, $I=1.2\times10^{14}$ W/cm$^{2}$, duration $40\,\mathrm{fs}$,
and wavelength $800\,\mathrm{nm}$; Boltzmann temperature 200 K.}
\end{figure}

In Fig. \ref{fig:04} we present the results of full calculations
for O$_{2}$, using Eq. (\ref{OxyHhgSig}), for the three geometries,
$\alpha=0^{o},\,45^{o}$, and $90^{o}$. The signals are seen to be
characterized by a full revival at $T_{rev}=\frac{1}{2Bc}=11.6\,\mathrm{ps}$
and also by the fractional $\frac{1}{2}$ and $\frac{1}{4}$ revivals,
like in N$_{2}$, as well as an additional $\frac{1}{8}$-revival,
for all the three geometries; the same characteristics have been observed
experimentally (e.g. \cite{kan-01,miy-02}). The existence of the
$\frac{1}{8}$-revival is due mainly to the presence of higher powers
and moments than $\left\langle \left\langle \cos^{2}\theta\right\rangle \right\rangle $,
that couple the Raman-forbidden ($\Delta J=\pm4$) and the {}``anomalous''
transitions ($|\Delta J|>4$) between the rotational states \cite{fai_rou-01}.
We may express the contribution from the first term of Eq. (\ref{OxyHhgSig})
more explicitly as: \begin{eqnarray}
S\left(t_{d};\alpha\right) & \approx & \frac{q_{1}}{64}\left\langle \left\{ \left(3-30\cos^{2}\alpha+35\cos^{4}\alpha\right)\right.\vphantom{\left\{ \left(30\cos^{2}\alpha\right)\right.^{2}}\right.\nonumber \\
 &  & \times\left\langle \sin^{2}\theta\cos^{2}\theta\right\rangle \nonumber \\
 &  & -\left(1-6\cos^{2}\alpha+5\cos^{4}\alpha\right)\left\langle \cos^{2}\theta\right\rangle \nonumber \\
 &  & +\left.\left.\left(4\sin^{2}\alpha-3\sin^{4}\alpha\right)\right\} ^{2}\right\rangle +\cdots.\label{OxySigApp}\end{eqnarray}
 For $\alpha=0^{o}$, this gives, $S\left(t_{d};\alpha\right)\approx q_{1}\left\langle \left\langle \sin^{2}\theta\cos^{2}\theta\right\rangle ^{2}\right\rangle $
and, for $\alpha=90^{o}$, $S\left(t_{d};\alpha\right)\approx q_{1}(\frac{3}{8})^{2}\left\langle \left\{ \left\langle \sin^{2}\theta\cos^{2}\theta\right\rangle -\frac{1}{3}\left\langle \cos^{2}\theta\right\rangle +\frac{1}{3}\right\} ^{2}\right\rangle $.
A comparison of the above expressions suggests that for $\alpha=0^{o}$
and 90$^{o}$, the signals at the full, $\frac{1}{2}$ and $\frac{1}{4}$
revivals would be in opposite phase, and that at the $\frac{1}{8}$
revival would be in phase. A direct comparison of the calculations
using the above abbreviated formulas with the full calculation in
Fig. 4 and the experimental observations in O$_{2}$ (e.g. \cite{kan-01,miy-02}),
fully confirm the above expectations. %
\begin{figure}
\begin{centering}
\includegraphics[scale=0.3]{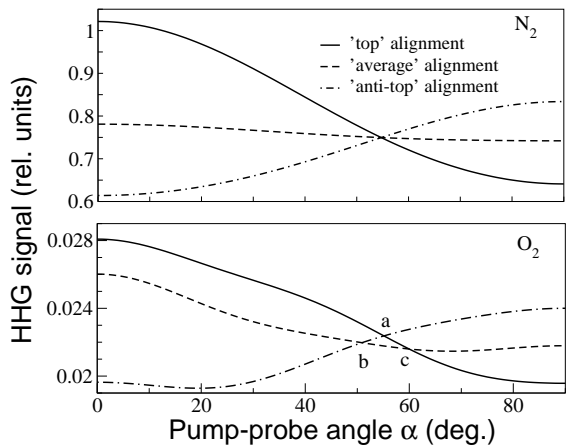} 
\par\end{centering}

\caption{\label{fig:05} Variation of dynamic HHG signal as a function of
pump-probe angle $\alpha$, near the first half-revival, for $N_{2}$
and $O_{2}$. The pulse parameters are the same as in Fig. (\ref{fig:03})
for $N_{2}$ and Fig. (\ref{fig:04}) for $O_{2}$.}
\end{figure}

In Fig. \ref{fig:05} we show the calculated results of the dynamical
signals for N$_{2}$ (upper panel) and O$_{2}$ (lower panel), as
a continuous function of $\alpha$, between 0$^{o}$ to 90$^{o}$,
at three different delay-times $t_{d}$ near the $\frac{1}{2}$ revival
period. In the upper panel for N$_{2}$, a remarkable coincidence
of the three signals is seen to occur at the {}``magic angle'' $\alpha_{c}=\arcsin\sqrt{\frac{2}{3}}\approx54.7^{o}$
as predicted above. Moreover, the signal at the {}``top''-alignment
time $t_{d}=4.05\,\mathrm{ps}$ (solid curve) is seen to lie above
the signal at the {}``anti-top'' alignment time $t_{d}=4.3\,\mathrm{ps}$
(dash-dot curve), for all $\alpha<\alpha_{c}$, and they invert their
relative strengths for all $\alpha>\alpha_{c}$. This is again in
agreement with the recent observations (e.g. \cite{ita-02,miy-01}).
The corresponding signals for O$_{2}$ (lower panel) does not show
a single crossing point, rather they cross at three different points
a, b, and c in the neighborhood of the magic angle $\alpha_{c}\approx54.7^{o}$.
Such a crossover-neighborhood around the magic angle $\approx54.7^{o}$
for O$_{2}$ is recently confirmed experimentally \cite{miy-02}.
The absence of a single crossing point for O$_{2}$ is due mainly
to the non-negligible contribution of the moment $\left\langle \left\langle \cos^{4}\theta\right\rangle \right\rangle $
to the O$_{2}$ signal (cf. Eq. (\ref{OxySigApp})).

Before concluding, we may make a few qualitative remarks on the $\alpha$
dependence of the dynamic signals for the more complex triatomic molecule
CO$_{2}$ \cite{miy-02}, and the organic molecule acetylene ($\pi$
symmetry), that are mesured recently \cite{kaj-02,tor-01}. The structure
of the operator (\ref{HhgOpe}) shows, even without a detailed calculation,
that the CO$_{2}$ and acetylene (HC$\equiv$CH), due to their linear
structure, would show a similar crossover at or near $\alpha_{c}\approx54.7^{o}$.
A direct perusal of the experimental data \cite{miy-02,kaj-02} confirms
this general expectation from the present theory -- both CO$_{2}$
and acetylene exhibit the crossover effect, and indeed near the {}``magic
angle'' $\alpha_{c}=\arcsin\sqrt{\frac{2}{3}}\approx54.7^{o}$.

To summarize: The simultaneous dependence of the dynamic HHG signals
from coherently rotating linear molecules, on the relative polarization
angle, $\alpha$, and the time delay, $t_{d}$, between a pump and
a probe pulse, is investigated theoretically. A general formula for
the dynamic signals for linear molecules is derived (Eqs. (\ref{Sig})
and (\ref{HhgOpe})). It is used to analyze the recently observed
characteristics of the HHG signals from N$_{2}$ and O$_{2}$. Among
other things, a {}``magic angle'' $\alpha_{c}=\arcsin\sqrt{\frac{2}{3}}\approx54.7^{o}$
for the crossing of the dynamic signals for N$_{2}$, and a crossover-neighborhood
around the {}``magic angle'', for O$_{2}$, are predicted by the
theory and confirmed by the available experimental data. The presence
of analogous crossovers for the more complex linear molecules, CO$_{2}$,
HC$\equiv$CH (acetylene), are also suggested by the present theory,
and are corroborated by the recent observations.

\begin{acknowledgments}
We thank Prof. K. Miyazaki for the private communications and fruitful
discussions. This work was partially supported by NSF through a grant
for ITAMP at Harvard University and Smithsonian Astrophysical Observatory. 
\end{acknowledgments}

\end{document}